\documentclass[twocolumn,showpacs,preprintnumbers,aps,a4]{revtex4}
\usepackage{epsfig}
\def\beq{\begin{eqnarray}}
\def\eeq{\end{eqnarray}}

\begin{document}

\preprint{q-bio/XXXXXXX}

\title{Cooperative action in eukaryotic gene regulation:
physical properties of a viral example}
\author{Maria Werner}

\author{LiZhe Zhu}

\author{Erik Aurell}
\affiliation{
Dept. Computational Biology\\
KTH -- Royal Institute of Technology\\
AlbaNova University Center, SE-106 91~Stockholm, Sweden
}%

\date{\today}

\begin{abstract}
The Epstein-Barr virus (EBV) infects more than 90\% of the human population, and is the
cause of several both serious and mild diseases. It is a tumorivirus, and
has been widely studied as a model system for gene (de)regulation in human. 
A central feature of the EBV life cycle is its ability to persist in 
human B cells in states denoted latency I, II and III. In latency III the host cell is
driven to cell proliferation and hence expansion of the viral population, but 
does not enter the lytic pathway, and no new virions are produced, while the latency I
state is almost completely dormant. In this paper we study a
physico-chemical model of the switch between latency I and latency III in EBV.
We show that the unusually large number of binding sites of two competing
transcription factors, one viral and one from the host, serves to make
the switch sharper (higher Hill coefficient), either by cooperative binding
between molecules of the same species when they bind, or by competition between
the two species if there is sufficient steric hindrance. 
\end{abstract}


\maketitle

\section{Introduction}
\label{s:introduction}
Genetic switches, mainly in bacteria, have recently interested statistical physicists, and work in this direction has been extensively reviewed in \cite{bintu,bintu2}. The fundamental assumption is that gene transcription, the copying of a stretch of DNA into mRNA, is either ``on'' or ``off''. This state of transcription depend on whether certain gene specific DNA binding proteins, transcription factors, are bound, or not, to the promoter region of the gene. A gene may be controlled by one or more transcription factors, each having a varying number of binding sites in the promoter region. The action of the transcription factor may in turn be either inhibitory or excitatory. Inhibition can arise from blocking access of the RNA-Polymerase to the transcription start site, while a stimulating effect is obtained if the bound factor stabilizes the Polymerase-DNA complex. A paradigmatic example where both effects are present is lysogeny maintenance in phage $\lambda$\cite{Ptashne,PtashneGann}. DNA looping, where distantly bound transcription factors interact and affect transcription, is also possible.

At a given transcription factor concentration, each possible state of promoter bound factors occurs with a probability given by a grand canonical ensemble formula.  The promoter region with the binding sites (with or without transcription factors) corresponds to the small system, and the cytoplasm, with a large number of transcription factors moving around, serves as the reservoir. Quite often transcription factors bind in dimer (or multimer) form, in which case the relevant concentration is determined by balance from the total concentration. In summary, the rate of transcription is a non-linear, sometimes quite complicated, function of the concentrations of the transcription factors regulating the gene.

One important property in gene regulation is cooperativity. If a single copy of a protein molecule in monomer form were to (positively) regulate a certain gene, the activity of that gene would follow the well-known Michaelis-Menten curve. The transcription rate would then be proportional to the concentration of the regulating molecule, up to a threshold above which it would level off. In other words, there would be appreciably high transcription even at very low concentrations of the regulating protein. The rationale for transcription factor often binding in multimer form, and of multiple DNA binding sites enabling cooperative interactions, is therefore assumed to be that it results in a sharper, more ``all-or-nothing'' switch.

Multiple binding sites for one and the same transcription factor are common in eukaryotic promoters. The object of this paper is one particular viral example of no less than 20 binding sites for a viral factor, where transcriptional activity has been observed to require 8 bound molecules \cite{wysokenski,zetterberg}, see section~\ref{s:Cp-EBV} below.  In addition, these sites are interleaved with an equal number of binding sites of a host transcription factor, presumably imposing the opposite effect. In a previous contribution,~\cite{werner}, we introduced, for reasons of computational simplicity, a thermodynamic model of this promoter switch ignoring eventual cooperative bindings and allowing some steric hindrance. Although direct experimental evidence is lacking, cooperative bindings of the viral transcription factor at this promoter is likely to be present, as well as more extensive blocking scenarios due the closely spaced sites. Both these mechanisms are likely to affect the sharpness of the switch.

We show in this paper that while cooperative protein 
interactions is one way to achieve  
effective cooperativity of the switch, accounting for full 
steric hindrance (blocking) of one species of molecules on the other 
is a more effective one. Therefore, 
a possible functional role of the alternating pattern 
of binding sites could be increasing effective cooperativity 
when the promoter architecture do not allow for cooperative molecular interactions.

The paper is organized as follows: in section~\ref{s:Cp-EBV} we describe our example, and in section~\ref{s:cooperativity-and-competition} we describe our model of cooperativity and competition in this example. In section~\ref{s:results} we summarize and discuss our results.

\section{The Epstein-Barr virus, the EBNA-1 protein, and the C promoter}
\label{s:Cp-EBV}

The Epstein-Barr virus (EBV) belongs to the gamma-herpes virus family, with relatives among other primate lymphocryptoviruses, and has likely co-evolved with man for a very long time~\cite{Gerner}. Although not discovered until the 1960ies, it is now known to infect more than 90 \% of the human population. The infection is asymptomatic if it occurs early in life, while later infection may result in infectious mononucleosis, more commonly known as ``the kissing disease''. The virus infects new hosts by virus particles shed from epithelial cells in the throat, and can persist in the host blood B cells for long times, in at least three distinct latent states known as latency I, II and III. EBV is medically important primarily because some cancer forms are invariably associated with the viral infection ~\cite{young2004}.

The most vital EBV protein is EBNA-1, a transcription factor involved in replication, episome partitioning as well as gene regulation ~\cite{leight}. In latency I, EBNA-1 is produced from RNA transcripts originating from the Q promoter on the EBV genome. EBNA-1 down-regulates transcription from Qp by binding to sites downstream of the transcription start site \cite{sample}. In latency III, on the other hand, EBNA-1 is produced together with five other proteins by alternative splicing of a longer RNA transcribed from the EBV C promoter (Cp) ~\cite{bodescot}. EBNA-1 positively regulates Cp activity by binding to the ``family-of-repeats'' (FR) region, positioned upstream of the start site \cite{reisman}. The physical description of this regulatory element is the topic of the present paper.

The FR region consists of 20 consecutive binding sites for EBNA-1~\cite{nilsson}. There are minor variations in the DNA sequence among these sites, but they are all experimentally verified, and approximately equally strong, binding sites~\cite{ambinder}. Comparing promoter activity, from constructs with varying number of binding sites in FR, revealed that at least eight sites are necessary to have full transcriptional activation \cite{wysokenski,zetterberg}, see Table~\ref{t:zetterberg}. Recent studies have identified an equal number of octamer binding sites at FR, juxtaposed with the EBNA-1 sites~\cite{almqvist}. The action of the human transcription factor Oct-2 , identified as binding to these octamer sites complex with the co-factors Groucho/TLE, is believed to be inhibitory~\cite{malin}.

In summary, the Cp activity is largely regulated by binding of two species of molecules, EBNA-1 and  Oct-2. They each can bind to 20 sites, and have antagonistic effects when bound. Due to the closely spaced binding sites, Oct-2 and EBNA-1 compete for binding to FR. It is however not experimentally known if one bound Oct-2 blocks out one or both of the neighbouring sites for EBNA-1, and \textit{vice versa}. The other unknown aspect is whether there exists cooperative binding between EBNA-1 proteins at FR, and if so, the strength of these interactions \cite{pLF}. Therefor we explore the effects of cooperative binding and blocking, with emphasis on how the effective cooperativity of the promoter switch is affected, \textit{i.e.} the sharpness of the switch.

\section{Cooperative binding and competition}
\label{s:cooperativity-and-competition}
The general thermodynamic framework is the following. Suppose a number of transcription factors $TF_1$, $TF_2$,\ldots, $TF_m$ can bind in different states indexed by $s$ around the start of a gene. The number of transcription factors of type $TF_i$ bound in state $s$ is $n_i(s)$, the association free energy is $\Delta G_s$, and the rate of transcription of the gene is $R_s$. Suppose further $[TF_i]$ is the concentration of transcription factor $TF_i$ in the surrounding cytoplasm, in the form in which this transcription factor binds. Then the binding sites, with or without bound transcription factors, can be considered a small system, exchanging particles (transcription factors) and energy with the larger reservoir. The probability of the small system being in state $s$ is

\begin{equation}
\label{eq:prob}
P_s \propto [TF_1]^{n_1(s)}\cdots [TF_m]^{n_m(s)}\exp(-\frac{\Delta G_s}{RT})  
\end{equation}
and the net average rate of transcription is
\begin{equation}
\label{eq:rate}
R([TF_1],\ldots,[TF_m]) = \sum_s R_s P_s
\end{equation}

The key assumption behind (\ref{eq:rate}) is that the time scale at which the probabilities in (\ref{eq:rate}) equilibrate is much faster than the time scales at which the the concentrations  $[TF_1]$, $[TF_2]$,\ldots, $[TF_m]$ change appreciably.
\\\\
In the present example, states can be labeled by $n$, the number of EBNA-1 molecules bound, $k$, the number of Oct-2 molecules bound, $n_1$, the number of cooperative bindings between bound EBNA-1 molecules and $k_1$,  the number of cooperative bindings between bound Oct-2 molecules. Every such state has a binding free energy of
\begin{equation}
\Delta G_{n,k,n_1,k_1} = n E_E + k E_O + n_1 E_{E1} + k_1 E_{O1}
\end{equation}
where  $E_E=-15.45$ kcal/mol \cite{ambinder} and $E_O=-12.28$ kcal/mol \cite{shah} are the known binding free energies of EBNA-1 and Oct-2 to binding sites in FR, and $E_{E1}$ and $E_{O1}$ are the unknown cooperative binding energies. In the numerical experiments described in this paper we only examine EBNA-1 cooperativity. $E_{E1}$ is proportional to $E_E$ in the range from 0 \% (no cooperativity) up to 40 \%. The total probability of the states with given values of $n$, $k$, $n_1$ and $k_1$ is hence 

\begin{equation}
P_{n,k,n_1,k_1} \propto \xi(n,k,n_1,k_1)[E]^n [O]^k \exp\left(-\frac{\Delta G_{n,k,n_1,k_1}}{k_B T}\right)
\end{equation}

where $\xi(n,k,n_1,k_1)$ is the number such states, and the overall rate of transcription is

\begin{equation}
\label{eq:sums}
P = \sum_{n=8}^{N} \sum_{k=0}^{N-n} 
\sum_{n_1=0}^{n-1} \sum_{k_1=0}^{N-n-1} P_{n,k,n_1,k_1}
\end{equation}

where $N$ is the number of binding sites. 

\begin{figure}[h]
\includegraphics[width = 0.45\textwidth]{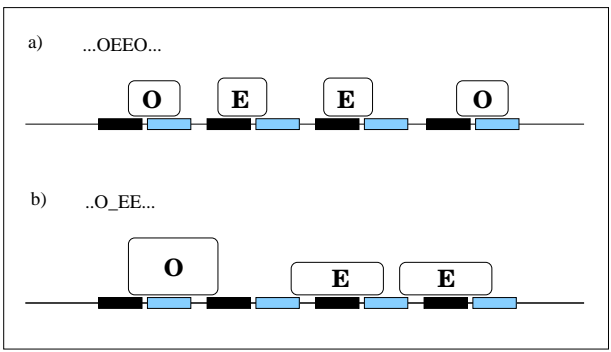}
\caption{ Illustration of the two blocking scenarios. EBNA-1 (E) bounds the black sites while Oct-2 (O) bound the blue sites. a) The single-side blocking model where a bound E blocks the closest O site to the right, and a bound O blocks E binding to the closest site on the left. The 40 binding sites are represented as 20 sites, where each site can be bound by E or O. b) The double-side blocking model where one bound E or O blocks the opposite molecular species binding on both sites. The 20 sites can be bound by E or O, with the restriction that there has to be an empty site between any bound O with a bound E on the right.}
\label{fig:blocking}
\end{figure}

As described briefly in the introduction, one can imagine two plausible blocking scenarios at FR. The first and simplest, is that each molecule bound hinders binding of the competitive species to the closest neighbouring site on one side. This is referred to a single-side blocking (Fig \ref{fig:blocking}a). The other scenario is that each bound molecule, sterically hinders both neighbouring sites for the other molecule; a double-sided blocking (Fig \ref{fig:blocking}b). The blocking method naturally affects the number of possible bound configurations, seen in Eq. \ref{eq:sums}. The upper bound in the sum over $k$ is $N-n$ in the single-side blocking model, but at most $N-n-1$ in the double-side blocking model for all $n$ greater than zero. Similarly, the sums over $n_1$ and $k_1$ may effectively go over smaller ranges \textit{e.g.} in the double blocking scenario with both molecules bound and $n+k=N-1$ all EBNA-1 and Oct-2 molecules bind together in two groups, hence $n_1=n-1$ and $k_1=k-1$. 

 Brute-force counting of $\xi(n,k,n_1,k_1)$ is not feasible as the number of states in this model is up to $3^{20}\approx 3.4 \cdot 10^{9}$ (in the model with single-side blocking only). Efficient calculation of  $\xi(n,k,n_1,k_1)$ involves two aspects. First, elementary combinatorics is used to build up a paradigm ``balls-baskets'' problem. It counts, under different constrains, the number of ways that one can put certain number of balls into another number of baskets. Second, we find a way that can describe efficiently all effects including double-side blocking, cooperativity and combination of both in a three-step algorithm:

\begin{enumerate}
\item 
Construct a backbone sequence (S0) made up by two types of baskets ($b_E$, $b_O$), the two types of molecules.
\item 
Distribute $n$ Es and $k$ Os among these baskets, forming a sequence (S1) consisting only of E and O.
\item 
Consider the front, end and the $n+k-1$ in-between positions of S1 as baskets ($b_\phi$) for empty binding sites $\phi$. Insert $N-(n+k)$ empty sites into these positions and get the final pattern (S2).
\end{enumerate}

By setting $N=20$, the actual number of sites is reduced by half, and the single-sided blocking model is the default. The double-side blocking is realized by setting the $b_\phi$ between an ''OE'' segment in S2 as must-be-filled baskets (Fig 1b). The number of cooperative units, $n_1$,  are counted by recording number of ''EE'' in S2, minus the number of $b_\phi$ that have been filled with $\phi$.
\\\\
To examine the effective cooperativity in the transition from $P\approx 0$ to $P\approx 1$ we compute the Hill coefficient. This is the logarithmic derivative of the ratio of probability of transcription to the probability of no transcription, with respect to the logarithm of the free ligand concentration. The Hill coefficient is a function of the ligand concentration, but the effective Hill coefficient is customarily taken at half saturation:

\begin{equation}
\frac {d\lg\frac{P}{1-P}} {d\lg[E_{free}]}\quad \mbox{at}\quad P=0.5
\end{equation}

In this paper we explore the Hill coefficient functions to see how blocking and cooperative binding influence the effective cooperativity of the switch. There are three cases studied; 1)~cooperative binding of EBNA-1 and no competing molecular species, 2)~cooperative binding of EBNA-1 with single-side blocking between the competing species, and 3)~cooperative binding of EBNA-1 with double-side blocking between the competing species. 

\section{Effective cooperativity of the switch}
\label{s:results}
\begin{figure}
\includegraphics[width = 0.5\textwidth]{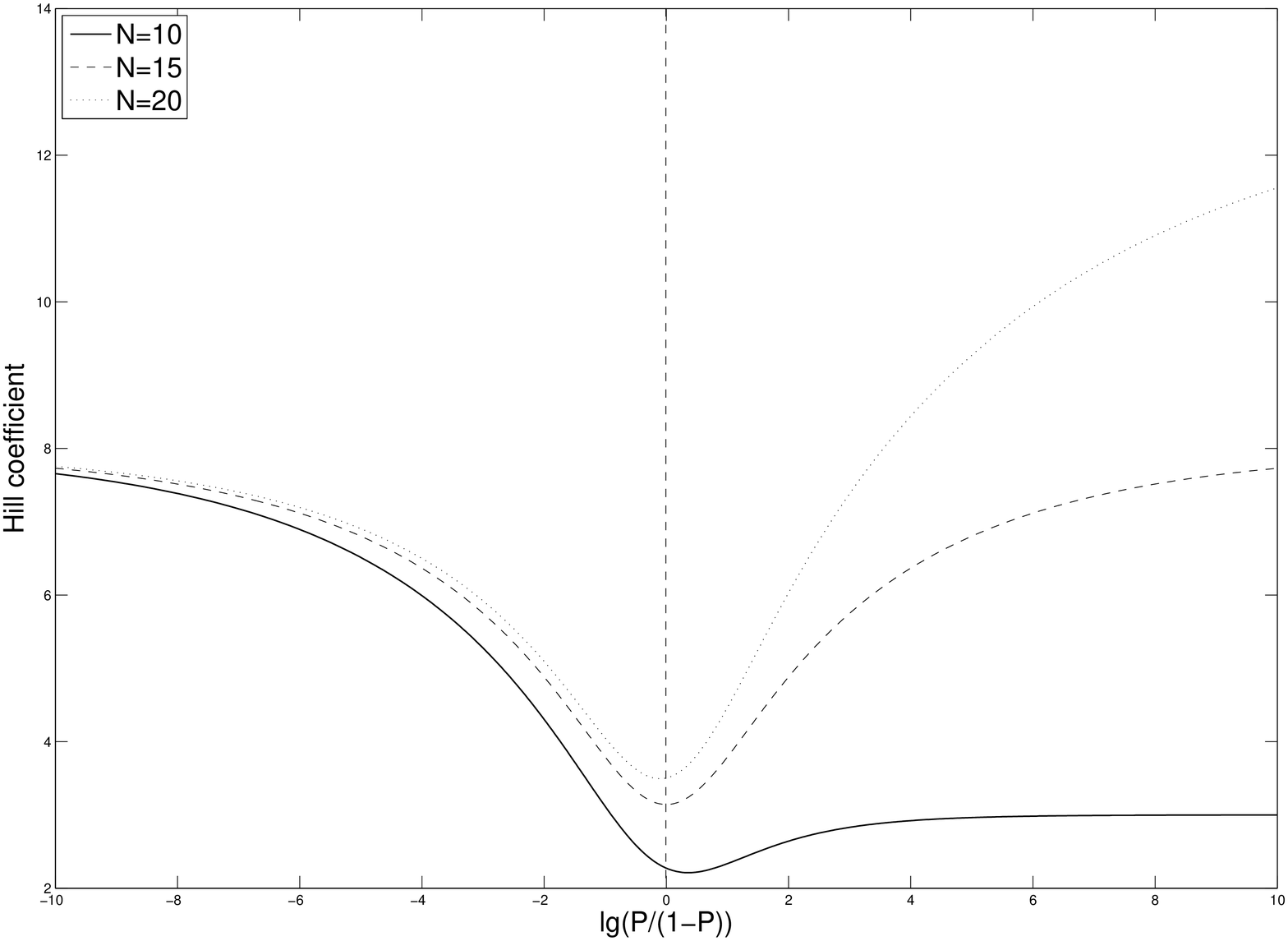}
\caption{Hill coefficient curves for three different numbers of binding sites in the system; 10, 15 or 20. For each case, the limit at low EBNA-1 concentrations (low P) is at Hill coefficient 8, since this is the definition of when the transcription is on at minimum. For the upper limit, at high P, the Hill coefficient approaches $N-7$.}
\label{fig:Nsites}
\end{figure}

One convenient way to visualize the cooperativity of the switch is as the ratio $\frac{P}{1-P}$ \textit{vs.} the local Hill coefficient given as $\frac {d\lg\frac{P}{1-P}} {d\lg[E_{free}]}$. For very high and very low concentrations of EBNA-1, corresponding to very large and very small values of $P$, it is easy to see that in our model $1-P\sim A [E_{free}]^{7-N}$ respectively $P\sim B[E_{free}]^{8}$. $A$ and $B$ are constants, and $N$ is the total number of binding sites in FR. Accordingly, the extreme local Hill coefficients are $N-7$ and $8$. Fig.~\ref{fig:Nsites} illustrates this limit behaviour for three values of $N$. 

\begin{figure}
\includegraphics[width = 0.5\textwidth]{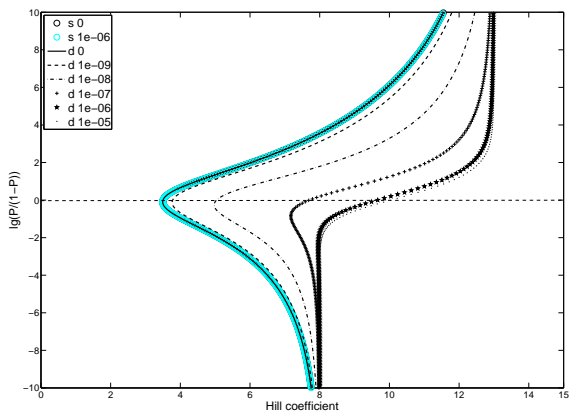}
\caption{Hill coefficient curves for both single- and double-side blocking case. The circles correspond to the single-side blocking, where the Oct-2 concentration do not at all affect the effective cooperativity. On the other hand, in the double-sided blocking model, the Oct concentration dramatically alters the cooperativity of the switch. For saturating levels of Oct-2, the effective Hill coefficient approaches 10.5 (dotted black line).}
\label{fig:Octlevels}\end{figure}

In the region of main interest, where $P\sim\frac{1}{2}$, the Hill coefficient curves show very different behavior for the three models. Without any cooperative interactions, and without competition, the effective Hill coefficient is substantially lower than both its limits. This baseline function for the system has an effective Hill coefficient of 3.5 (Fig \ref{fig:Octlevels}, circled lines). This low Hill coefficient remains even with competition from Oct-2 binding, for the single-side blocking, the effective cooperativity practically insensitive to Oct-2 levels.  On the contrary, competition with double-side blocking dramatically alters the shape of the Hill coefficient curve, to a sigmoidal interpolation between the limits $8$ and $N-7$. The effective Hill coefficient then changes from 3.5 up to 10.5, for saturating amounts of Oct-2 (Fig \ref{fig:Octlevels}, dotted lines). \\

From a theoretical point of view, the thermodynamic model of the switch is a (finite, one-dimensional) Ising-like model with three states at each site: bound by EBNA-1, bound by Oct-2, or free. The only complication in computing the ``ON'' probability ($P$) is that only states with enough bound EBNA-1 count, which mixes in a global variable in the elementary statistical mechanical model. The single-blocking results can however be readily understood. With no cooperative binding and only single blocking, one can sum over $k$ in (\ref{eq:werner-probs}) to obtain the model studied in~\cite{werner}, that is 

\begin{equation}
P_{n} \propto \left({N \atop n}\right)[E]^n e^{-n\frac{E_E}{k_BT}}
\left(1+[O]e^{-\frac{E_O}{k_B T}}\right)^{N-n}
\label{eq:werner-probs}
\end{equation}

Including the normalization this means

\begin{equation}
P_{n} = \left({N \atop n}\right)\frac{z^n}{(1+z)^N}
\qquad z=\frac{[E]e^{\frac{E_E}{k_BT}}}{\left(1+[O]e^{-\frac{E_O}{k_B T}}\right)}
\label{eq:werner-probs-2}
\end{equation}

and the ratio between ON and OFF probabilities is therefore a function of the variable $z$ only: 

\begin{equation}
\frac{P}{1-P} = \frac{\sum_{n=8}^{20}\left({N \atop n}\right)z^n}
                     {\sum_{n=0}^{7}\left({N \atop n}\right)z^n} = f(z)
\label{eq:werner-probs-3}
\end{equation}

The local Hill coefficients are

\begin{equation}
\frac {d\lg\frac{P}{1-P}} {d\lg[E_{free}]}= \frac {d\lg f(z)} {d\lg z} 
\label{eq:werner-probs-4}
\end{equation}

which like the ratio $\frac{P}{1-P}$ depends on the concentration of the second molecule $[O]$ only through $z$. The effective cooperativity in the model without  cooperative binding and only single blocking hence does not depend on  $[O]$, as shown in the curves in Fig~\ref{fig:Octlevels}. The Hill coefficient at $P\approx \frac{1}{2}$ can be estimated by approximating the binomials with a Gaussian distribution, \textit{i.e.}
\begin{equation}
\label{eq:werner-probs-5}
P \approx  C^{-1}\int_{\bar{x} + c\sigma}^{\infty}
        \exp\left(-\frac{1}{2\sigma^2}(x-\bar{x})^2 + x\log z\right)dx
\end{equation}
where, in the case at hand, 
$C = e^{\frac{\sigma^2}{2}\log^2 z}\sqrt{\frac{2\pi}{\sigma^2}}$,
$\bar{x}=\frac{N}{2}=10$, $\sigma=\sqrt{\frac{N}{4}}=\sqrt{5}$
and  $c=-\frac{2}{\sigma}$. Half-filling is achieved at $z^*=e^{c/\sigma}$,
and the Hill coefficient is $\sqrt{8\sigma^2/\pi}\approx 3.6$
which accords quite well with the minimum value in Fig.~\ref{fig:Octlevels}.
The switch is therefore much less sharp than the limits of $8$ and $N-7=13$, at respectively $P\approx 0$ and  $P\approx 1$ could have led one to believe. We note that the sharpness increases with $N$
(as long as the threshold stays around $N/2$), but only as the square root of $N$: more than
a hundred consecutive binding sites are necessary to reach a Hill coefficient of about ten
in a model of this kind.
\\\\
In the model with double blocking on the other hand clearly the effective cooperativity
can be much larger, and also depend on $[O]$. That is easy to understand in the limit where $[O]$ is large; if so EBNA-1 and Oct-2 compete for binding sites, and the possibility that a site is left free can be disregarded. Therefore, if $n$ copies of EBNA-1 are bound, then also $N-n$ copies of Oct-2 are bound, altogether in the pattern $EEEE\cdots OOOO$ with statistical weight

\begin{equation}
P_{n} = \frac{x^n}{1+x+x^2+\cdots x^N}
\qquad x=\frac{[E]e^{\frac{E_E}{k_BT}}}{[O]e^{-\frac{E_O}{k_B T}}}
\label{eq:double-blocking-high-O}
\end{equation} 

The Hill coefficient is then only a function of $x$, such that the curve in Fig~\ref{fig:Octlevels} has a limit when $[O]$ becomes large, and the value of the Hill coefficient at \textit{e.g.} $x=1$ then lies between the limits of $8$ and $13$. Competition with a second molecule therefore makes the switch sharper for double-side blocking, in contrast to the situation in single-sided blocking.\\

The case with cooperativity can be understood qualitatively, with the helix-coil model of protein physics. Without Oct-2, the statistical model can be written as a factor $h=[E]e^{-\frac{E_E+E_{coop}}{k_BT}}$ for each letter $E$, and a penalty $c=e^{\frac{E_{coop}}{k_BT}}$ for every start letter of a string of $E$'s. In an infinitely long string, the \textit{fraction} of letters $E$ as well as the \textit{frequency of initiation} of a string of $E$'s are calculated from the leading eigenvalue of the transfer matrix~\cite{SneppenZocchi}. In our case, the interesting region is obviously when that fraction is around $40\%$, as $8$ sites out of $20$ need to be filled to have transcription from Cp. If $c$ is close to one, cooperative binding is weak, and the switch is similar to the single-blocking case discussed above. If on the other hand $c$ is 
much less than one, the expected fraction of letters $E$ can be larger than $40\%$, while the expected frequency if initialization of a string of $E$'s is less than once in twenty sites. Eventually, we would expect that either all twenty sites are bound, or no sites in FR be bound. 
This describes a situation where all twenty molecules have to bind in simultaneously,
in which case the Hill coefficient is 20.

\begin{figure}[h]
\includegraphics[width = 0.5\textwidth]{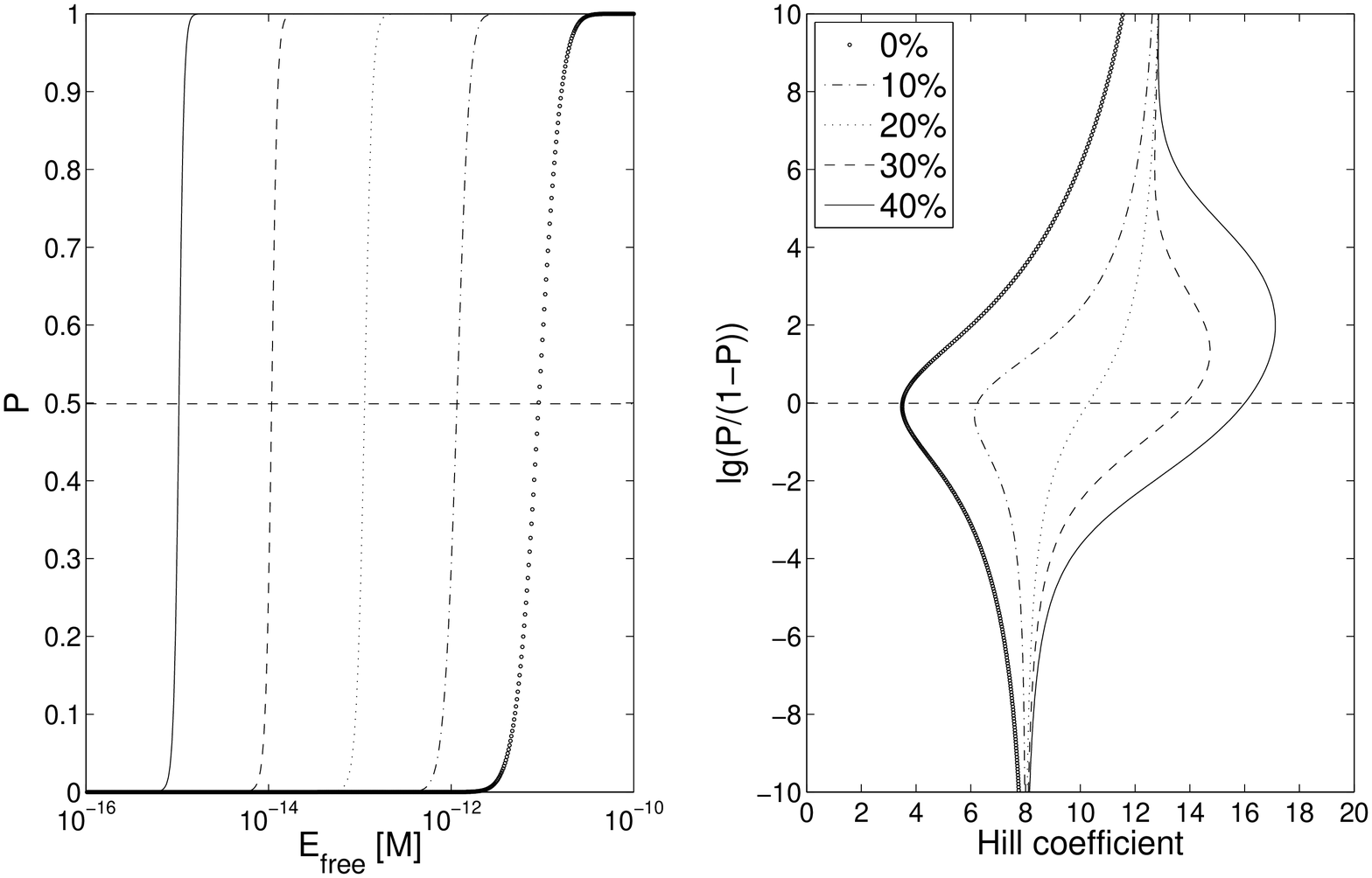}
\caption{Hill coefficient curves for the model without any competitive molecular species (Oct-2) but with increasing strength of cooperative binding between EBNA-1. The cooperative binding is varied from 0-40 \% of the DNA binding strength of EBNA-1. This range corresponds to 0-6.2 kcal/mol in binding energies. With no competition and only added cooperative interaction, the effective Hill coefficient changes dramatically from 3.5 up to 16 for the 6.2 kcal/mol cooperative binding energy.}
\label{fig:NoOct}\end{figure}

The addition of a cooperative binding of EBNA-1 to both the single- and double sided model, hence changes the effective Hill coefficient differently, depending on model. Fig \ref{fig:NoOct} displays the curves for 5 different cooperative binding strengths, when no Oct-2 is competing for the FR sites. The range of cooperative strength here is from 0 \% up to 40 \% of the DNA affinity, \textit{i.e} $\approx 6.2 $ kcal/mol, where the effective Hill coefficient is increased from 3.5 to 16. 

\begin{figure}[h]
\includegraphics[width = 0.5\textwidth]{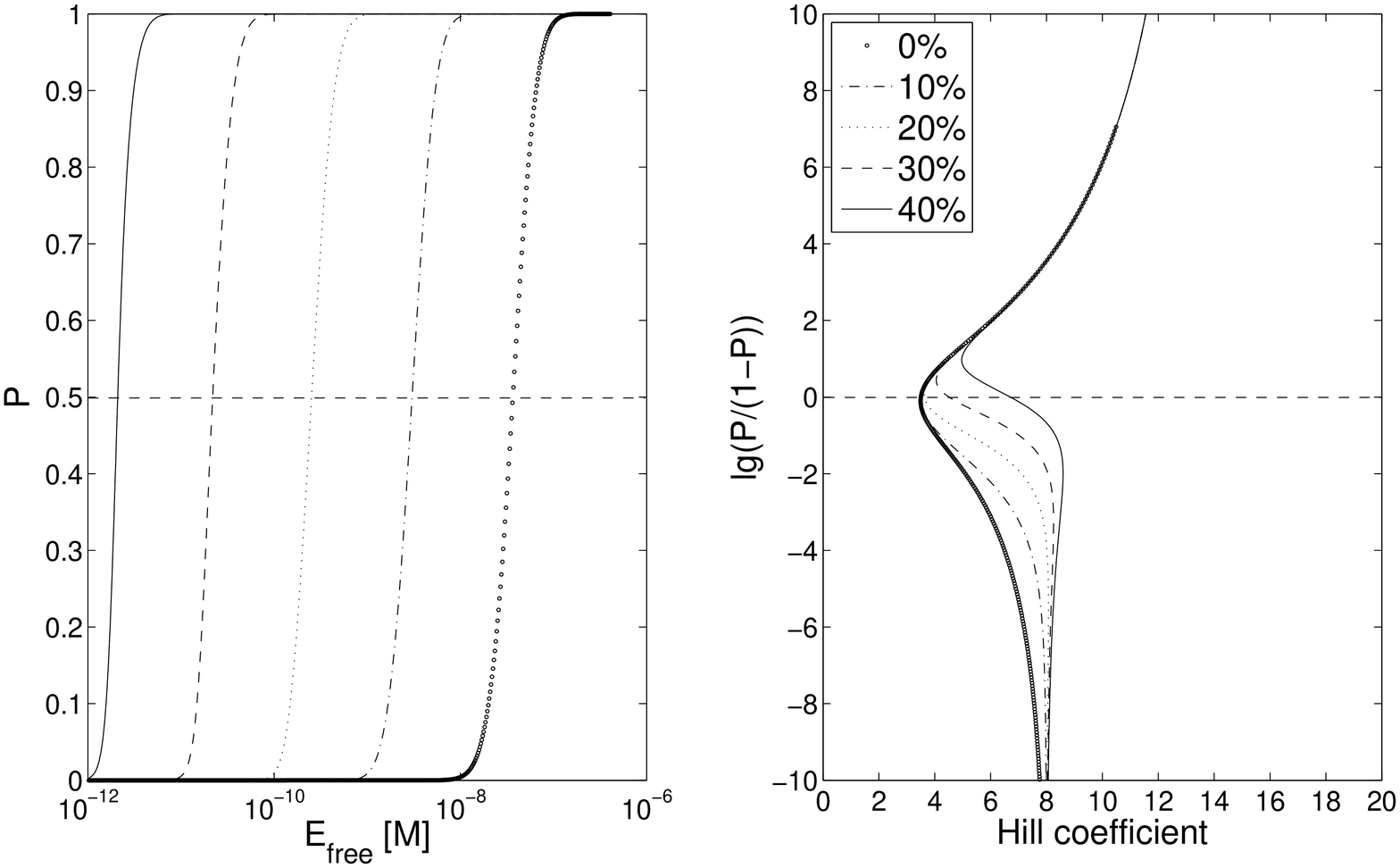}
\caption{Hill coefficient curves for the model with high concentration of the competitive molecular species (Oct), single-side blocking and different strength of cooperative binding between EBNA-1. The cooperative binding is varied from 0-40 \% of the DNA binding strength of EBNA-1. The effect of adding cooperative bindings for EBNA-1 only increases the effective Hill coefficient from 3.5 to 7.}
\label{fig:Highsingle}
\end{figure}

However, in the real system the competitive protein Oct-2 is likely to be present, perhaps even at very high concentrations. As for the single-sided blocking, an additional cooperative binding of EBNA-1 does not have the same impact when Oct-2 levels are high. Instead of a 4-fold change, from 3.4 to 16, the effective Hill coefficient is now only doubled, from 3.5 to 7 (compare Fig \ref{fig:NoOct} and \ref{fig:Highsingle}, solid lines). This is to be compared with the double-sided blocking model, where even no cooperative bindings have a relatively high effective cooperativity. Adding up to 40 \% cooperative binding strength, the Hill coefficient is almost doubled, from 10.5 to 18 (Fig \ref{fig:Octlevels}) \\

A conclusion to draw from this is that to create an effective switch for genetic control, this type of architecture, with alternating binding sites for two antagonistic factors, can be one approach. For EBV, the FR region is known for its enhancer function, as well as forming a looped structure with another EBNA-1 binding region on the viral genome; the dyad symmetry (DS) ~\cite{frappier1991b,su}. This structure is involved in replication initiation control. If the EBNA-1 binding sites in FR were to be arrange in the same manner as in DS, \textit{i.e.} much closer in space, there might be cooperative bindings forming even at FR. However, since FR also seem to play an important role in forming a looped structure, there might be a structural reason behind these more sparsely placed sites, not enabling the same type of tight interactions. And, as we show here, there is no need for cooperative interactions to get a sharp switch of Cp activity, as long as there is efficient steric hindrance.

\begin{figure}[h]
\includegraphics[width = 0.5\textwidth]{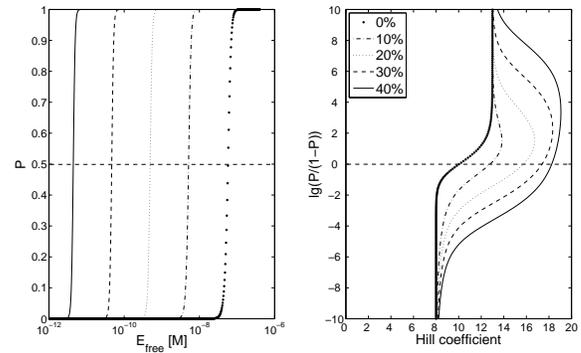}
\caption{Hill coefficient curves for the model with high concentration of the competitive molecular species (Oct), double-side blocking and different strength of cooperative binding between EBNA-1. The cooperative binding is varied from 0-40 \% of the DNA binding strength of EBNA-1. Double blocking in itself, gives a high effective Hill coefficient, and the extra cooperative interactions almost doubles this coefficient up to 18.}
\label{fig:Highdouble}
\end{figure}

\acknowledgements{We thank Ingemar Ernberg for sharing his knowledge of the Epstein-Barr virus
and many discussions on mechanisms and modelling of the EBV lat I/lat III switch. 
This work was supported by the Swedish Science Council (M.W. and E.A.). }

\bibliography{references}

\begin{thebibliography}{22}
\expandafter\ifx\csname natexlab\endcsname\relax\def\natexlab#1{#1}\fi
\expandafter\ifx\csname bibnamefont\endcsname\relax
  \def\bibnamefont#1{#1}\fi
\expandafter\ifx\csname bibfnamefont\endcsname\relax
  \def\bibfnamefont#1{#1}\fi
\expandafter\ifx\csname citenamefont\endcsname\relax
  \def\citenamefont#1{#1}\fi
\expandafter\ifx\csname url\endcsname\relax
  \def\url#1{\texttt{#1}}\fi
\expandafter\ifx\csname urlprefix\endcsname\relax\def\urlprefix{URL }\fi
\providecommand{\bibinfo}[2]{#2}
\providecommand{\eprint}[2][]{\url{#2}}

\bibitem[{\citenamefont{Bintu et~al.}(2005{\natexlab{a}})\citenamefont{Bintu,
  Buchler, Garcia, Gerland, Hwa, Kondev, Kuhlman, and Phillips}}]{bintu}
\bibinfo{author}{\bibfnamefont{L.}~\bibnamefont{Bintu}},
  \bibinfo{author}{\bibfnamefont{N.~E.} \bibnamefont{Buchler}},
  \bibinfo{author}{\bibfnamefont{H.~G.} \bibnamefont{Garcia}},
  \bibinfo{author}{\bibfnamefont{U.}~\bibnamefont{Gerland}},
  \bibinfo{author}{\bibfnamefont{T.}~\bibnamefont{Hwa}},
  \bibinfo{author}{\bibfnamefont{J.}~\bibnamefont{Kondev}},
  \bibinfo{author}{\bibfnamefont{T.}~\bibnamefont{Kuhlman}}, \bibnamefont{and}
  \bibinfo{author}{\bibfnamefont{R.}~\bibnamefont{Phillips}},
  \bibinfo{journal}{Curr Opin Genet Dev} \textbf{\bibinfo{volume}{15}},
  \bibinfo{pages}{116} (\bibinfo{year}{2005}{\natexlab{a}}).

\bibitem[{\citenamefont{Bintu et~al.}(2005{\natexlab{b}})\citenamefont{Bintu,
  Buchler, Garcia, Gerland, Hwa, Kondev, Kuhlman, and Phillips}}]{bintu2}
\bibinfo{author}{\bibfnamefont{L.}~\bibnamefont{Bintu}},
  \bibinfo{author}{\bibfnamefont{N.~E.} \bibnamefont{Buchler}},
  \bibinfo{author}{\bibfnamefont{H.~G.} \bibnamefont{Garcia}},
  \bibinfo{author}{\bibfnamefont{U.}~\bibnamefont{Gerland}},
  \bibinfo{author}{\bibfnamefont{T.}~\bibnamefont{Hwa}},
  \bibinfo{author}{\bibfnamefont{J.}~\bibnamefont{Kondev}},
  \bibinfo{author}{\bibfnamefont{T.}~\bibnamefont{Kuhlman}}, \bibnamefont{and}
  \bibinfo{author}{\bibfnamefont{R.}~\bibnamefont{Phillips}},
  \bibinfo{journal}{Curr Opin Genet Dev} \textbf{\bibinfo{volume}{15}},
  \bibinfo{pages}{125} (\bibinfo{year}{2005}{\natexlab{b}}).

\bibitem[{\citenamefont{Ptashne}(2005)}]{Ptashne}
\bibinfo{author}{\bibfnamefont{M.}~\bibnamefont{Ptashne}},
  \emph{\bibinfo{title}{A genetic switch (3rd edition)}}
  (\bibinfo{publisher}{Cold Spring Harbour Laboratory Press},
  \bibinfo{year}{2005}).

\bibitem[{\citenamefont{Ptashne and Gann}(2002)}]{PtashneGann}
\bibinfo{author}{\bibfnamefont{M.}~\bibnamefont{Ptashne}} \bibnamefont{and}
  \bibinfo{author}{\bibfnamefont{A.}~\bibnamefont{Gann}},
  \emph{\bibinfo{title}{Genes and Signals}} (\bibinfo{publisher}{Cold Spring
  Harbour Laboratory Press}, \bibinfo{year}{2002}).

\bibitem[{\citenamefont{Wysokenski and Yates}(1989)}]{wysokenski}
\bibinfo{author}{\bibfnamefont{D.~A.} \bibnamefont{Wysokenski}}
  \bibnamefont{and} \bibinfo{author}{\bibfnamefont{J.~L.} \bibnamefont{Yates}},
  \bibinfo{journal}{Journal of Virology} \textbf{\bibinfo{volume}{63}},
  \bibinfo{pages}{2657} (\bibinfo{year}{1989}).

\bibitem[{\citenamefont{Zetterberg et~al.}(2004)\citenamefont{Zetterberg,
  Borestrom, Nilsson, and Rymo}}]{zetterberg}
\bibinfo{author}{\bibfnamefont{H.}~\bibnamefont{Zetterberg}},
  \bibinfo{author}{\bibfnamefont{C.}~\bibnamefont{Borestrom}},
  \bibinfo{author}{\bibfnamefont{T.}~\bibnamefont{Nilsson}}, \bibnamefont{and}
  \bibinfo{author}{\bibfnamefont{L.}~\bibnamefont{Rymo}},
  \bibinfo{journal}{International Journal of Oncolocy}
  \textbf{\bibinfo{volume}{25}}, \bibinfo{pages}{693} (\bibinfo{year}{2004}).

\bibitem[{\citenamefont{Werner et~al.}()\citenamefont{Werner, Ernberg, Zou,
  Almqvist, and Aurell}}]{werner}
\bibinfo{author}{\bibfnamefont{M.}~\bibnamefont{Werner}},
  \bibinfo{author}{\bibfnamefont{I.}~\bibnamefont{Ernberg}},
  \bibinfo{author}{\bibfnamefont{J.}~\bibnamefont{Zou}},
  \bibinfo{author}{\bibfnamefont{J.}~\bibnamefont{Almqvist}}, \bibnamefont{and}
  \bibinfo{author}{\bibfnamefont{E.}~\bibnamefont{Aurell}},
  \bibinfo{note}{submitted to BMC Systems Biology}.

\bibitem[{\citenamefont{Gerner et~al.}(2004)\citenamefont{Gerner, dolan, and
  McGeoch}}]{Gerner}
\bibinfo{author}{\bibfnamefont{C.}~\bibnamefont{Gerner}},
  \bibinfo{author}{\bibfnamefont{A.}~\bibnamefont{dolan}}, \bibnamefont{and}
  \bibinfo{author}{\bibfnamefont{D.}~\bibnamefont{McGeoch}},
  \bibinfo{journal}{Virus Research} \textbf{\bibinfo{volume}{99}},
  \bibinfo{pages}{187} (\bibinfo{year}{2004}).

\bibitem[{\citenamefont{Young and Rickinson}(2004)}]{young2004}
\bibinfo{author}{\bibfnamefont{L.~S.} \bibnamefont{Young}} \bibnamefont{and}
  \bibinfo{author}{\bibfnamefont{A.~B.} \bibnamefont{Rickinson}},
  \bibinfo{journal}{Nat Rev Cancer} \textbf{\bibinfo{volume}{4}},
  \bibinfo{pages}{757} (\bibinfo{year}{2004}).

\bibitem[{\citenamefont{Leight and Sugden}(2000)}]{leight}
\bibinfo{author}{\bibfnamefont{E.~R.} \bibnamefont{Leight}} \bibnamefont{and}
  \bibinfo{author}{\bibfnamefont{B.}~\bibnamefont{Sugden}},
  \bibinfo{journal}{Reviews in Medical Virology} \textbf{\bibinfo{volume}{10}},
  \bibinfo{pages}{83} (\bibinfo{year}{2000}).

\bibitem[{\citenamefont{Sample et~al.}(1992)\citenamefont{Sample, Henson, and
  Sample}}]{sample}
\bibinfo{author}{\bibfnamefont{J.}~\bibnamefont{Sample}},
  \bibinfo{author}{\bibfnamefont{E.~B.} \bibnamefont{Henson}},
  \bibnamefont{and} \bibinfo{author}{\bibfnamefont{C.}~\bibnamefont{Sample}},
  \bibinfo{journal}{J Virol} \textbf{\bibinfo{volume}{66}},
  \bibinfo{pages}{4654} (\bibinfo{year}{1992}).

\bibitem[{\citenamefont{Bodescot et~al.}(1987)\citenamefont{Bodescot,
  Perricaudet, and Farrell}}]{bodescot}
\bibinfo{author}{\bibfnamefont{M.}~\bibnamefont{Bodescot}},
  \bibinfo{author}{\bibfnamefont{M.}~\bibnamefont{Perricaudet}},
  \bibnamefont{and} \bibinfo{author}{\bibfnamefont{P.~J.}
  \bibnamefont{Farrell}}, \bibinfo{journal}{Journal of Virology}
  \textbf{\bibinfo{volume}{61}}, \bibinfo{pages}{3424} (\bibinfo{year}{1987}).

\bibitem[{\citenamefont{Reisman and Sugden}(1986)}]{reisman}
\bibinfo{author}{\bibfnamefont{D.}~\bibnamefont{Reisman}} \bibnamefont{and}
  \bibinfo{author}{\bibfnamefont{B.}~\bibnamefont{Sugden}},
  \bibinfo{journal}{Molecular and cellular Biology}
  \textbf{\bibinfo{volume}{6}}, \bibinfo{pages}{3838} (\bibinfo{year}{1986}).

\bibitem[{\citenamefont{Nilsson et~al.}(2001)\citenamefont{Nilsson, Zetterberg,
  Wang, and Rymo}}]{nilsson}
\bibinfo{author}{\bibfnamefont{T.}~\bibnamefont{Nilsson}},
  \bibinfo{author}{\bibfnamefont{H.}~\bibnamefont{Zetterberg}},
  \bibinfo{author}{\bibfnamefont{Y.~C.} \bibnamefont{Wang}}, \bibnamefont{and}
  \bibinfo{author}{\bibfnamefont{L.}~\bibnamefont{Rymo}},
  \bibinfo{journal}{Journal of Virology} \textbf{\bibinfo{volume}{75}},
  \bibinfo{pages}{5796} (\bibinfo{year}{2001}).

\bibitem[{\citenamefont{Ambinder et~al.}(1990)\citenamefont{Ambinder, Shah,
  Rawlins, Hayward, and Hayward}}]{ambinder}
\bibinfo{author}{\bibfnamefont{R.~F.} \bibnamefont{Ambinder}},
  \bibinfo{author}{\bibfnamefont{W.~A.} \bibnamefont{Shah}},
  \bibinfo{author}{\bibfnamefont{D.~R.} \bibnamefont{Rawlins}},
  \bibinfo{author}{\bibfnamefont{G.~S.} \bibnamefont{Hayward}},
  \bibnamefont{and} \bibinfo{author}{\bibfnamefont{S.~D.}
  \bibnamefont{Hayward}}, \bibinfo{journal}{Journal of Virology}
  \textbf{\bibinfo{volume}{64}}, \bibinfo{pages}{2369} (\bibinfo{year}{1990}).

\bibitem[{\citenamefont{Almqvist et~al.}(in press)\citenamefont{Almqvist, Zou,
  Linderson, Borestrom, Altiok, Zetterberg, Rymo, Petterson, and
  Ernberg}}]{almqvist}
\bibinfo{author}{\bibfnamefont{J.}~\bibnamefont{Almqvist}},
  \bibinfo{author}{\bibfnamefont{J.}~\bibnamefont{Zou}},
  \bibinfo{author}{\bibfnamefont{Y.}~\bibnamefont{Linderson}},
  \bibinfo{author}{\bibfnamefont{C.}~\bibnamefont{Borestrom}},
  \bibinfo{author}{\bibfnamefont{E.}~\bibnamefont{Altiok}},
  \bibinfo{author}{\bibfnamefont{H.}~\bibnamefont{Zetterberg}},
  \bibinfo{author}{\bibfnamefont{L.}~\bibnamefont{Rymo}},
  \bibinfo{author}{\bibfnamefont{S.}~\bibnamefont{Petterson}},
  \bibnamefont{and} \bibinfo{author}{\bibfnamefont{I.}~\bibnamefont{Ernberg}},
  \bibinfo{journal}{Journal of General Virology} pp.~\bibinfo{pages}{--}
  (\bibinfo{year}{in press}).

\bibitem[{\citenamefont{Malin et~al.}(2005)\citenamefont{Malin, Linderson,
  Almqvist, Ernberg, Tallone, and Petterson}}]{malin}
\bibinfo{author}{\bibfnamefont{S.}~\bibnamefont{Malin}},
  \bibinfo{author}{\bibfnamefont{Y.}~\bibnamefont{Linderson}},
  \bibinfo{author}{\bibfnamefont{J.}~\bibnamefont{Almqvist}},
  \bibinfo{author}{\bibfnamefont{I.}~\bibnamefont{Ernberg}},
  \bibinfo{author}{\bibfnamefont{T.}~\bibnamefont{Tallone}}, \bibnamefont{and}
  \bibinfo{author}{\bibfnamefont{S.}~\bibnamefont{Petterson}},
  \bibinfo{journal}{Nucleic Acids Research} \textbf{\bibinfo{volume}{33}},
  \bibinfo{pages}{4618} (\bibinfo{year}{2005}).

\bibitem[{\citenamefont{Frappier}()}]{pLF}
\bibinfo{author}{\bibfnamefont{L.}~\bibnamefont{Frappier}},
  \bibinfo{note}{personal communication to M. Werner}.

\bibitem[{\citenamefont{Shah et~al.}(1997)\citenamefont{Shah, Bertolino, and
  Singh}}]{shah}
\bibinfo{author}{\bibfnamefont{P.~C.} \bibnamefont{Shah}},
  \bibinfo{author}{\bibfnamefont{E.}~\bibnamefont{Bertolino}},
  \bibnamefont{and} \bibinfo{author}{\bibfnamefont{H.}~\bibnamefont{Singh}},
  \bibinfo{journal}{The EMBO Journal} \textbf{\bibinfo{volume}{16}},
  \bibinfo{pages}{7105} (\bibinfo{year}{1997}).

\bibitem[{\citenamefont{Sneppen and Zocchi}(2005)}]{SneppenZocchi}
\bibinfo{author}{\bibfnamefont{K.}~\bibnamefont{Sneppen}} \bibnamefont{and}
  \bibinfo{author}{\bibfnamefont{G.}~\bibnamefont{Zocchi}},
  \emph{\bibinfo{title}{Physics in Molecular Biology}}
  (\bibinfo{publisher}{Cambridge University Press}, \bibinfo{year}{2005}).

\bibitem[{\citenamefont{Frappier and O'Donnell}(1991)}]{frappier1991b}
\bibinfo{author}{\bibfnamefont{L.}~\bibnamefont{Frappier}} \bibnamefont{and}
  \bibinfo{author}{\bibfnamefont{M.}~\bibnamefont{O'Donnell}},
  \bibinfo{journal}{Proc.Natl. Acad. Sci} \textbf{\bibinfo{volume}{88}},
  \bibinfo{pages}{10875} (\bibinfo{year}{1991}).

\bibitem[{\citenamefont{Su et~al.}(1991)\citenamefont{Su, Middelton, Sudgen,
  and Echols}}]{su}
\bibinfo{author}{\bibfnamefont{W.}~\bibnamefont{Su}},
  \bibinfo{author}{\bibfnamefont{T.}~\bibnamefont{Middelton}},
  \bibinfo{author}{\bibfnamefont{B.}~\bibnamefont{Sudgen}}, \bibnamefont{and}
  \bibinfo{author}{\bibfnamefont{H.}~\bibnamefont{Echols}},
  \bibinfo{journal}{Proc. Natl. Acad. Sci.} \textbf{\bibinfo{volume}{88}},
  \bibinfo{pages}{10870} (\bibinfo{year}{1991}).

\end{thebibliography}

\newpage
\section*{Tables}
\begin{table}
\label{t:zetterberg}
\begin{tabular}{|l|l|} \hline
{\bf Number of sites} & {\bf Activity} \\ \hline
20 & 280 \\
19 & 229 \\
17 & 226 \\
14 & 169 \\
12 & 206 \\
11 & 169 \\
8 &  87\\
6 &  19\\
5 &  19\\
4 &  11\\
3 &  3.3\\
2 &  2.1\\
1 &  1.2\\
0 &  3.3\\ \hline
\end{tabular}
\caption{Activity of Cp promotor in EBV strains with
different numbers of binding sites for EBNA-1 in
the family of repeats site, adapted after
\protect\cite{zetterberg}. Activity level
relative to control.}
\end{table}

\end{document}